# A Python Framework for Fast Modelling and Simulation of Cellular Nonlinear Networks and other Finite-difference Time-domain Systems

Radu Dogaru
dept. of Applied Electronics and Information Engineering
University "Politehnica" of Bucharest
Bucharest, Romania
radu.dogaru@upb.ro

Ioana Dogaru
dept. of Applied Electronics and Information Engin
University "Politehnica" of Bucharest
Bucharest, Romania
ioana.dogaru@upb.ro

*Abstract*—This paper introduces and evaluates a freely available cellular nonlinear network simulator optimized for the effective use of GPUs, to achieve fast modelling and simulations. Its relevance is demonstrated for several applications in nonlinear complex dynamical systems, such as slow-growth phenomena as well as for various image processing applications such as edge detection. The simulator is designed as a Jupyter notebook written in Python and functionally tested and optimized to run on the freely available cloud platform Google Collaboratory. Although the simulator, in its actual form, is designed to model the FitzHugh Nagumo Reaction-Diffusion cellular nonlinear network, it can be easily adapted for any other type of finite-difference time-domain model. Four implementation versions are considered, namely using the PyCUDA, NUMBA respectively CUPY libraries (all three supporting GPU computations) as well as a NUMPY-based implementation to be used when GPU is not available. The specificities and performances for each of the four implementations are analyzed concluding that the PyCUDA implementation ensures a very good performance being capable to run up to 14000 Mega cells per seconds (each cell referring to the basic nonlinear dynamic system composing the cellular nonlinear network). (*Abstract*)

## I. Introduction

In fractal analysis and nonlinear dynamics, the use of cellular nonlinear networks (CNNs) [1] and more generally finite-difference time-domain (FDTD) models [2] is widespread. For instance, highly relevant phenomena such as Turing patterns, Spiral waves, other slow growth phenomena, tumor modelling, wave propagation and many others may be studied in cellular nonlinear networks frameworks. Particularly interesting, the Reaction-Diffusion CNNs exhibit various emergent dynamics phenomena and life-like behaviors [3][4]. For such networks, a viable theory for locating emergent behaviors in the parameter space (or gene's space) called local activity theory [5] was proposed and successfully tested [4]. Fluid dynamics, sound propagation, and many other physical phenomena can be modeled in FDTD frameworks such as cellular automata and Lattice Boltzmann Machines [6][7][8]. Such models need convenient informatic implementations (modelling and simulation frameworks = MSF), and in recent years various commercial or non-commercial solutions were offered, most struggling to offer GPU support and high performance (short simulation times for wide arrays of cells). For instance [9] is one of the most interesting non-commercial solutions for RD-CNNs but for fast simulations requires the use of a personal platform (PC) equipped with a graphical processing unit (GPU). On the other hand, the concept of reproducible research [10] gains more and more interest among research communities and it basically relies on some instruments, often cloud computing platforms, where one can run the same code and verify results as well as adding her/his own contributions. For instance, one can develop Jupyter notebooks [11] capable to run in a free or commercial cloud platform. The MSF discussed herein is thought under the reproducible research perspective and offers four implementation variants for a RD-CNN model that can be easily adapted to a more general FDTD model. Section II describes the mathematical model and how it is translated into 4 different implementations, all capable to run with GPU support freely offered via Google Collaboratory platform [12]. Section III gives an in-depth analysis of the dynamic performances (simulation times) and specific issues for each of the 4 implementations (based on PyCUDA, NUMBA, CUPY and NUMPY libraries) and for some of the GPU processors available on the cloud platform. Several applications including fast search of the parameter space in order to identify meaningful dynamics are given in section IV, with an emphasis on slow growth phenomena (needing good speed to observe dynamics in large arrays i.e. 4096x4096 cells for 100000 iterations) and meaningful image processing tasks such as edge detection. Concluding remarks are given in Section V, summarizing that our tool offers interesting perspectives for open research in nonlinear complex systems. Other models can be adapted straightforwardly, for instance the Cellular Neural Network (CeNN) [13] which, in the context of emerging artificial intelligence (AI) applications can be regarded as a recurrent convolutional neural network (CNN). Following guidelines in [14] and [15] where a NUMBA implementation was first proposed, and using the proposed MSF as a template, one can easily adapt it for the CNN or other models of interest.

## II. THE RD-CNN MODEL AND ITS FOUR IMPLEMENTATIONS

### A. The mathematical model and the general framework for the modelling and simulation process

The following discrete time model is considered in this paper (here exemplified for the FitzHugh Nagumo model) [14]. The equations defining the RD-CNN model are:

$$u_{i,j} = k_a x_{i,j}; \quad v_{i,j} = k_a x_{i,j};$$
$$\text{for } t = 1,...T \text{ and for all cells } (i,j)$$
$$\begin{bmatrix} u^+_{i,j} = u_{i,j} + \Delta t \left[ f_1(u_{i,j}, v_{i,j}) + D_u(u_{i+1,j} + u_{i-1,j} + u_{i,j-1} + u_{i,j+1} - 4u_{i,j}) \right] \\ v^+_{i,j} = v_{i,j} + \Delta t \left[ f_2(u_{i,j}, v_{i,j}) + D_v(v_{i+1,j} + v_{i-1,j} + v_{i,j-1} + v_{i,j+1} - 4v_{i,j}) \right] \\ u_{i,j} = u^+_{i,j} \\ v_{i,j} = v^+_{i,j} \end{bmatrix} \quad (1)$$

The two layers of the RD-CNN are given by the $u_{i,j}$ and $v_{i,j}$ state variables, corresponding to the nonlinear cells. The pair $(i,j) \in \{1,..NN, 1,..NM\}$ represents the spatial index of the cell located in a 2D array of $NNxNM$ size. The entire group of cells in the grid are associated to arrays **A** (comprising *u* values*)* and **B** (comprising *v* values). Initial state values can be programmed with some specific image files (typ=3) or with some randomly generated arrays (typ=2 – where all cells are randomly generated; typ=1 – where only a 11x11 square in the middle of the array has random values). For external images, a scaling factor $k_a$ may be considered. $D_u$ and $D_v$ are the diffusion coefficients associated to the layers. The nonlinear functions for the above model are given next, and they include some of the gene parameters $[c, a, b, eps]$:

$$f_1(u,v) = cu - \frac{1}{3}u^3 - v \quad (2)$$

$$f_2(u,v) = -eps(u - bv + a) \quad (3)$$

The entire set of the RD-CNN model parameters is called a *gene* and will be allocated to a single variable **Pars** as shown next. In the next subsection details on implementing the above model using 4 different approaches are given. For all these implementations there is a unique function **get_initial_state()** to read the parameters, construct the initial state arrays and the display buffers (A_show, B_show) and another function for displaying the result. These functions are defined in "CELL1" of the notebook which can be run only once. Next figure displays the arguments and returns of the input function which should be always called before running the model. One can identify the parameters discussed above.

```
6 def get_init_state(typ=1, img_size=512, name='test_pattern_1__.jpg',
7                    a=-0.3, b= 1.3, eps= -0.1, c=1, Du=0.06, Dv=1, dt=0.1, ka=1):
  ....
59     return (Pars, A,B,Anew,Bnew,A_show,B_show,colmap)
```

If another FDTD or cellular nonlinear model is considered a new specific function should be defined using the above as a starting template.

Another function **disp_simul()**, necessary to display and save the simulation result is shown in the next figure with its argument and returns (latest values of the **A, B** arrays after running the model *iter_max* iterations). The *nnsp* argument is an integer representing the number of snapshots to be displayed (for instance *nnsp*=5 snapshots, these images being stored into the **A_show** and **B_show** tensors). The last argument *implement* is a string which displays the type of simulator (one of the four to be described next) and it will be displayed on the simulation plot as shown in Figure 1.

```
61 def disp_simul(A_show,B_show,nsnp,iter_max, colmap, implement):
   ......
102    return(lastA,lastB)
```

The result of this function is an inline plot but a .png file is also generated, for further use, as shown in Fig.1 following a simulation using the PYCUDA library. The complete list of parameters is displayed as well as the dynamics of the A and B arrays during the specified simulation time.

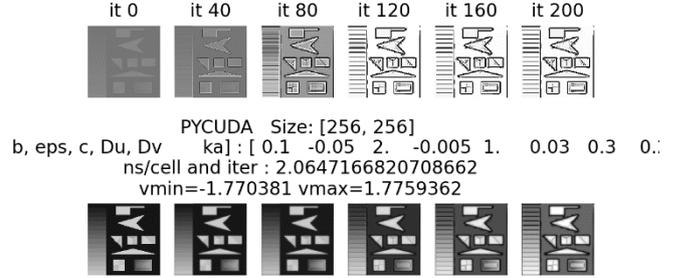

Fig. 1. The result of running a PyCUDA simulation

The dynamic performance is expressed here in "ns/cell and iteration" which is the reverse of the Mega-cells/second. In the above case 2.06ns correspond to 1000/2.06 = 484 Mega-cells/second. Running the next cell in the notebook [16] allows visualization of the latest **A** (or **B**) array (in the above case, the one after 200 iterations), as seen in Figure 2, in this case representing some edge extraction example (from the initial state figure displayed in iteration 0).

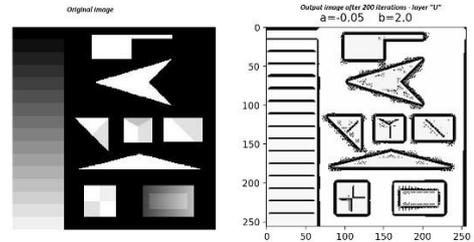

Fig. 2. The image associated with the A array after 200 iterations (right) is a meaningful transformation (gray level edge detection) of the initial state image on the left side.

The simulator is available in [16] and is constructed as a Jupyter notebook composed of o series of cells. The first three cells should be run only once, the first reporting on the specific GPU available in the current Google Collab runtime (one can try to get access to a better GPU using the "Factory reset runtime" in the "Runtime" tab. The second cell allows to install the PyCUDA library on Google Collab platforms. Running it may be not necessary on other platforms, where PyCUDA is by default installed (e.g. on Kaggle [17] platforms). The third cell implements the functions discussed previously to input the model parameters and display the results. Then, four cells each implementing the simulator using a different library can be used to effectively run the simulation. A brief description for each one is described next while the detailed implementation is available in [16].

## B. PyCUDA implementation

The PyCUDA library [18] can be regarded as a "wrapper" for the NVCC compiler, giving thus Pythonic access to the basic NVIDIA tools for programming GPUs. It offers simple functions to load and get variables on/from the GPU and then do computation on the GPU. Except CUDA kernel definition all remaining code is written in Python. The kernel definition for the model in Eq. (1)-(3) is given in Figure 3. As seen, it should follow the NVIDIA C syntax, since it is compiled by the NVIDIA **nvcc** compiler. The full code is available in [16], and some excerpt emphasizing the kernel definition is given in the next figure.

```
47    int n_x=$sizm;
48    int n_y=$sizn;
49    int i = threadIdx.x + blockDim.x*blockIdx.x;
50    int j = threadIdx.y + blockDim.y*blockIdx.y;
51
53    int i_left; int i_right; int j_down; int j_up;
55    // Boundary conditions
56    if(i==0) {i_left=n_y-1;} else {i_left=i-1;}
57    if(i==n_y-1) {i_right=0;} else {i_right=i+1;}
58    if(j==0) {j_down=n_x-1;} else {j_down=j-1;}
59    if(j==n_x-1) {j_up=0;} else {j_up=j+1;}
61    int cen,dre,stg,sus,jos;
62    cen=j+i*n_x; dre=j+n_x*i_right; stg=j+n_x*i_left; sus=j_up+n_x*i; jos=j_down+n_x*i;
64    Anew[cen] = A[cen]+Pars[0]*(A[cen]*(Pars[4]-A[cen]*Pars[3])-B[cen]+ Pars[5]*( A[dre] +A[stg] + A[jos] + A[sus] -4*A[cen]));
65    Bnew[cen]= B[cen]+Pars[0]*(-Pars[3]*(A[cen]-Pars[2]*B[cen]+Pars[1])+Pars[6]*(B[dre] +B[stg] + B[jos] +B[sus]  -4*B[cen]));
66
```

Fig. 3. The main part of the PyCUDA kernel deffinition associated to the RD-CNN model.

As seen, the model is entirely described in lines 64-65 after some useful definition of the positional indices in lines 61-62. The same positional definition given the localization of the thread in lines 49-50 can be reused when other cellular or FDTD models need to be implemented. The remaining of the code given in the PyCUDA cell does the following: i) reads the simulation parameters; ii) defines the CUDA **blockdim** and **griddim** sizes (here some optimization may improve the speed performance); iii) prepares the variables (**A, Anew, B, Bnew** and **Pars** arrays and stores them into the GPU global memory); iv) runs the simulation by calling the **ca_core()** function which executes the above kernels in parallel on the GPU while then swaps the **Anew** with **A** (and **B** with **Bnew**) preparing the process for the next iteration. Note that **Anew** represents the state of the **A** array after one iteration (when A was applied as initial state). All these specific steps are implemented, but in their specific manners, using the other 3 approaches (NUMBA, CUPY, NUMPY). The PYCUDA main loop is represented in the next figure:

```
104   for iter in range(iter_max):
105     # call of the kernel function
106     ca_core(A_gpu, B_gpu, Anew_gpu, Bnew_gpu, Pars_gpu, block=blockdim, grid=griddim)
107     # prepare for a new iteration
108     A_gpu = Anew_gpu
109     B_gpu = Bnew_gpu
110
111     # snapshots transffered from GPU to CPU
112     if (iter+1) % test_mod == 0:
113
114       print ("%5d, (elapsed: %f s)" % (iter, time.time()-timer))
115       A_show[1+iter//(test_mod),:,:]=Anew_gpu.get()
116       B_show[1+iter//(test_mod),:,:]=Bnew_gpu.get()
117
118   runtime = time.time() - timer
119   print (" total: %f s" % runtime)
```

Using the default parameters of the **get_init_state()** function, running the PYCUDA implementation will exhibit a slow growth phenomena as seen in Figure 4.

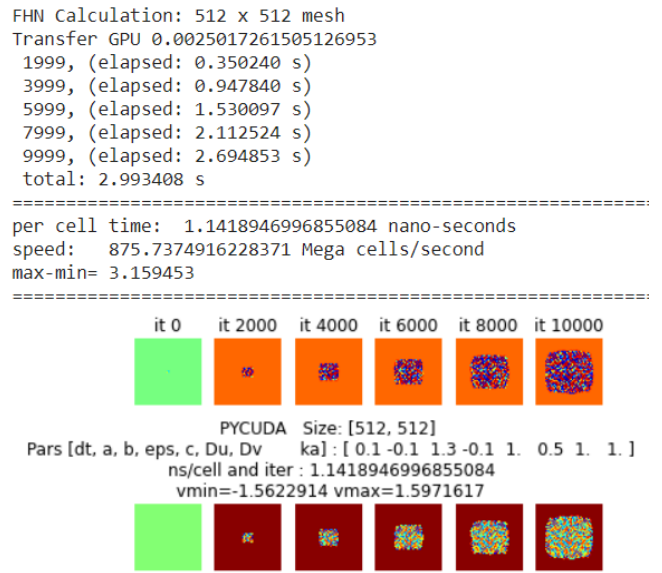

Fig. 4. Running simulation using the PYCUDA implementation.

As seen, only 2 seconds suffice to run the simulation of a 512x512 array during 10000 iterations. In this case the available GPU was not one of the best.

## C. NUMBA implementation

The NUMBA library [19] is intended to offer powerful compilers (JIT and CUDAJIT) for both GPU and CPU units. Unlike PyCUDA, the description of CUDA kernels is entirely done in Python, giving thus more portability. However, as seen in Section III, the GPU device is not as efficiently used as in the case of PYCUDA library. As in the case of the PYCUDA library, the cell used to implement the simulator has the same three components: i) reads simulation parameters; ii) defines CUDA block and grid dimensions; iii) prepares the specific variables and moves them on the GPU device; iv) runs the simulation by calling iteratively the GPU-executed and compiled function **ca_core()**. An excerpt of the code (fully provided in [16]) is given in the next figure, emphasizes the kernel definition.

```
31   @cuda.jit
32   def ca_core(A, B, Anew, Bnew, Pars):
33
34     n = A.shape[0]
35     m = A.shape[1]
36
37     # thread distribution of the cell implementation
38     j = cuda.threadIdx.x + cuda.blockIdx.x * cuda.blockDim.x
39     i = cuda.threadIdx.y + cuda.blockIdx.y * cuda.blockDim.y
40     if (j >= 0 and j < n) and (i >= 0 and i < m) :  # GPU equivalent of for loops in CPU
41
42         Anew[j, i] = A[j,i]+Pars[0]*(A[j,i]*(Pars[4]-A[j,i]*Pars[3])-B[j,i]+ \
43            Pars[5]*( A[j,(i+1)%m] +A[j,(i-1)%m] +A[(j+1)%n,i] -4*A[j,i]) )
44         Bnew[j,i]= B[j,i]+Pars[0]*(-Pars[3]*(A[j,i]-Pars[2]*B[j,i]+Pars[1])+ \
45            Pars[6]*(B[j,(i+1)%m] +B[j,(i-1)%m] + B[(j+1)%n,i] +B[(j+1)%n,i] -4*B[j,i]))
```

Fig. 5. Kernel definition for the NUMBA implementation of the RD-CNN model.

Unlike in the case of PyCUDA, 2-dimensional indices are directly used here, with the toroidal boundary condition implemented using the "modulo" operator in Python (%) applied to the positional indices. The entire model is implemented in lines 42-45 and is obviously easy to change it given another mathematical description of the FDTD model. As in the PyCUDA case, the simulation needs calling the kernel in a main loop and storing the snapshot arrays at some given moments. When running in the Google Collaboratory platforms, the first run on the NUMBA-based implementation may produce some error "Numba cannot operate on non-primary CUDA context …". In our experience, waiting for some time (up to 10 minutes) and repeating the run will end-up successfully. As seen in Table I in the next section the speed of the NUMBA implementation is worse, although there is an advantage of Python portability. On the other hand, installing NUMBA on various personal platforms is easier and less cumbersome than PYCUDA.

*D. NUMPY implementation*

In many circumstances, it is possible that a GPU device is not available. For such circumstances, a NUMPY – based implementation is considered as a separate cell. The excerpt from the code in [16] implementing the **ca_core()** function and its main loop is given in the next figure. NUMPY is a widely used library which implements efficient computing when arrays are used. Various linear algebra functions are efficiently implemented, thus allowing high performance without the need of an additional compiler (array methods are already pre-compiled).

The main model is now implemented in lines 18-21, and as seen, it is quite easy to replace it with a more general FDTD model. Also, it is quite straightforward to pass from one model implementation library to another, given the templates shown in Figures 3,5 and 6.

```
15 def ca_core_npy(A, B, Pars):
16
17    #x=Pars[0]*( A*(Pars[4]-A*A/3)
18    Anew = A+Pars[0]*(A*(Pars[4]-A*A/3)-B + \
19    Pars[5]*(np.roll(A,-1,0)+np.roll(A,1,0)+np.roll(A,-1,1)+np.roll(A,1,1)-4*A ))
20    Bnew= B+Pars[0]*( -Pars[3]*(A-Pars[2]*B+Pars[1])+ \
21    Pars[6]*(np.roll(B,-1,0)+np.roll(B,1,0)+np.roll(B,-1,1)+np.roll(B,1,1)-4*B ))
22    return (Anew, Bnew)
23 #   MAIN LOOP
24 print ( "FHN Calculation: %d x %d mesh" % (NN, NM))
25 timer = time.time()
26 for iter in range(iter_max):
27    # call one-step CA running
28    (A,B)=ca_core_npy(A, B, Pars)
29    # outputs become inputs
```

Fig. 6. NUMPY-based implementation of the ca_core() function implementing the RD-CNN model

It is important to note that here one operates with arrays directly (the specific NUMPY philosophy) while the neighborhoods and frontier conditions are implemented using the **np.roll()** method. A simulation using the same parameters as in Figure 4 is given in Figure 7. The result of the simulation is, as expected, similar but the simulation time is larger.

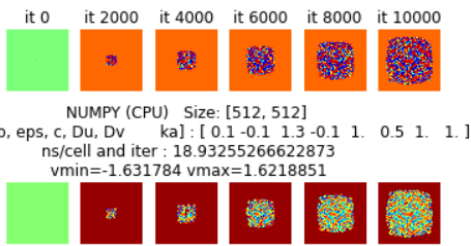

Fig. 7. The simulation of the RD-CNN model on CPU, using the NUMPY library.

As seen, the running time is smaller than in the case of GPU usage, however, when restricted platforms (CPU-only) are available using this implementation is a good option. As detailed in [15], CPU-s can be sometimes more effectively exploited using the JIT compiler from NUMBA library. The **ca_core()** model in this case is simply a rewrite of the GPU NUMBA model where @jit replaces the @cudajit and **for** loops would replace **if** directives.

*E. CUPY implementation of the RD-CNN model*

The CUPY library [20] was proposed as an alternative to NUMPY (many functions are similar) with GPU support. Consequently, code written for NUMPY can be readily transformed in CUPY-based code (with the advantage of GPU support) using some very simple rules: i) **np.method()** is replaced with **cp.method()** where **cp** is the alias for **cupy** as well as **np** is the alias for **numpy**; ii) in order to use GPUs, variables must be declared as GPU-variables and transferred among GPU and CPU using the specific methods: **Ag=cp.array(A)** – for copying the CPU array A into GPU array Ag, and **A=cp.asnumpy(Ag)** for the reverse operation. As a result, the **ca_core()** definition given in Figure 8 is very similar to the NUMPY definition, yet now the simulator benefits from GPU support, if GPU is available.

```
19 Ag=cp.array(A)
20 Bg=cp.array(B)
21 Anewg=cp.array(Anew)
22 Bnewg=cp.array(Bnew)
23 Parsg=cp.array(Pars)
24
25 def ca_core_cpy(Ag, Bg, Parsg):
26    Anewg = Ag+Parsg[0]*(Ag*(Parsg[4]-Ag*Ag/3)-Bg + \
27    Parsg[5]*(cp.roll(Ag,-1,0)+cp.roll(Ag,1,0)+cp.roll(Ag,-1,1)+cp.roll(Ag,1,1)-4*Ag ))
28    Bnewg= Bg+Parsg[0]*( -Parsg[3]*(Ag-Parsg[2]*Bg+Parsg[1])+ \
29    Parsg[6]*(cp.roll(Bg,-1,0)+cp.roll(Bg,1,0)+cp.roll(Bg,-1,1)+cp.roll(Bg,1,1)-4*Bg ))
30    Ag = Anewg
31    Bg = Bnewg
32    return (Ag,Bg)
33
34 #  Main loop
35 print ( "FHN Calculation: %d x %d mesh" % (NN, NM))
36 timer = time.time()
37 for iter in range(iter_max):
38    # One step iteration
39    (Ag,Bg)=ca_core_cpy(Ag, Bg, Parsg)
40    # Output becomes input (variables Ag Bg are stored on GPU's global memory )
41
```

Fig. 8. CUPY-based implementation of the RD-CNN model

Running the simulator for the same set of parameters gives, as expected, the same result, with some acceleration, but not as efficient as in the case of PYCUDA or NUMBA implementations.

*F. Performance comparison among implementations.*

In the following table a comparison is given between all the above-mentioned implementations. One important factor influencing the speed is the array size *N*. The GPU device was Tesla P100-PCIE and the CPU (NUMPY case) was Intel(R) Xeon(R) CPU @ 2.30GHz. Both devices were allocated by the Google Colab platform. It is likely that performance may slightly differ in other specific device allocation on the cloud platform. In order to compare performance with other simulators, the table reports the speed as Mcells/second i.e. representing how many CA cells (or in general FDTD cells) are computed in a second. The same model (parameters and running time 10000 iterations) as considered in the previously reported simulation was considered (it corresponds to the default **get_init_state()** function).

TABLE I. PERFORMANCE (MCELLS/SECOND) COMPARISON BETWEEN VARIOUS IMPLEMENTATIONS (SIMULATION TIME IS GIVEN IN PARENTHESIS)

|  | N=128 | N=256 | N=512 | N=1024 | N=2048 | N=4096 |
|---|---|---|---|---|---|---|
| **PYCUDA** | 470 (0,34) | 1941 (0.34) | 7581 (0.34) | 13059 (0.79) | 14198 (2.95) | 14545 (11.53) |
| **NUMBA** | 39 (4.18) | 151 (4.33) | 601 (4.36) | 2439 (4.29) | 4026 (10) | 4191 (40) |
| **CUPY** | 15.5 (10.5) | 42 (15.3) | 149 (17.4) | 720 (14.5) | 1308 (32) | 1353 (123) |
| **NUMPY (CPU)** | 26 (4.3) | 52 (12.5) | 53 (49) | 66 (158) | 44 (944) | — |

Clearly, that the most efficient GPU usage is given by the PYCUDA implementation, particularly when large arrays are considered the performance reaches up to 14000 Mcells/second. The NUMBA implementation, on the same GPU ensures a lower, 3-4 times less, speed. For small array size, using CPU can give satisfactory performance while CUPY becomes effective (with respect to NUMPY) for arrays with *N>512* but still not so effective as NUMBA or PYCUDA. In fact, this result demonstrates that for the problem considered here, namely the cellular nonlinear model implementation, the GPU-based linear algebra library supported by CUPY is not very well used when compared with user-defined CUDA kernel definition. Although CUPY offers the possibility for user-defined CUDA kernels, we did not implement it since PYCUDA already offers this approach and demonstrated a very good performance.

Next table indicates how computing speed is influenced by the choice of the GPU device (on Google Collab. platforms various GPUs are assigned upon their availability).

TABLE II. SPEED PERFORMANCE (MCELLS/SECOND) FOR VARIOUS GPU UNITS (THE CASE OF *N*=4096)

|  | Tesla K80 | Tesla T4 | Tesla P100-PCIE |
|---|---|---|---|
| **Speed (Mcells/seconds)** | 2519 | 12469 | 14545 |
| **Computing time (10000 iterations)** | 66 | 13 | 11.53 |

## III. APPLICATIONS

Having a fast CNN simulator opens interesting possibilities particularly when various complex dynamics behaviors should be associated with specific locations in the parameter space. In [4] an analytic method dubbed "edge of chaos" and based on local activity theory [5] was considered to roughly locate sets of parameters leading to meaningful and interesting dynamic behaviors. Although the local activity method is rather fast since it does not require RD-CNN simulation to detect the "edge of chaos profile", the theory does not include the diffusion coefficients and consequently more simulations are needed to finely locate the specific parameters leading to interesting dynamics. Such interesting dynamic phenomena are slow growth phenomena mimicking various natural phenomena including life, tumor formation and evolutions, pandemics etc. where large number of iterations (and consequently good simulator speeds) are a must. Other relevant dynamics is associated with image-processing such as edge extraction and other feature extractors. There is still a lot of un-explored potential for cellular nonlinear networks in artificial intelligence and we believe that using such fast simulators can reveal interesting applications. It is worth mentioning that a cellular nonlinear network acts as a convolutional layer with recurrence, i.e. applied on itself hundreds of times (as specified by the number of iterations) being thus equivalent to a very deep convolutional network (the depth being equivalent to the number of iterations). Such recurrent implementation is an economical alternative to the usual feed-forward convolution neural network, not mentioning that CeNNs (cellular nonlinear networks) are more plausible brain models. In the following we will exemplify with two examples of complex behaviors that can be easily modelled and simulated with the proposed code.

### A. Fast identfication of emergent behaviors (slow growth)

Starting from results in [14] one can use our simulator to rapidly explore some regions in the (*Du,Dv*) diffusion parameter space. As seen in Figure 9, a large variety of behaviors cand be detected using square random mid patters as initial state or fully random initial states (as on the rightmost row in Figure 9)

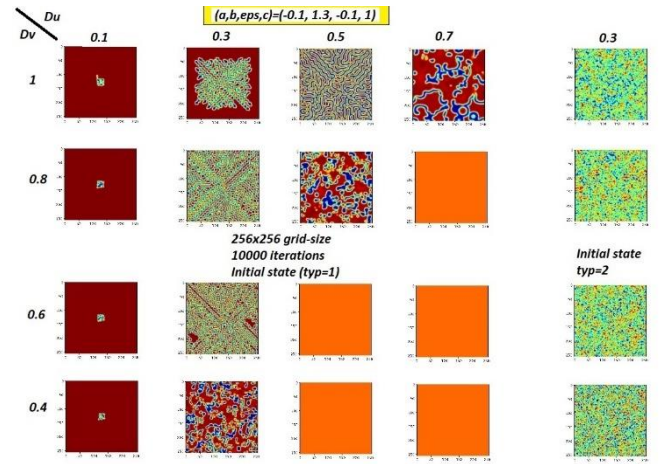

Fig. 9. Simulation of the RD-CNN model for various diffusion coefficients

A diversity of dynamic emergent phenomena can be revealed, ranging from slow-growth phenomena *(Du,Dv)=(0.3,1)* to dynamic patterns *(Du,Dv)=(0.5,0.8)* or dull dynamics (a homogeneous steady state) as in the case of *(Du,Dv)=(0.7,0.8)*. Figure 10 represents an example of slow growth while Figure 11 gives the final pattern (after 10000 iterations) in four random initial state cases.

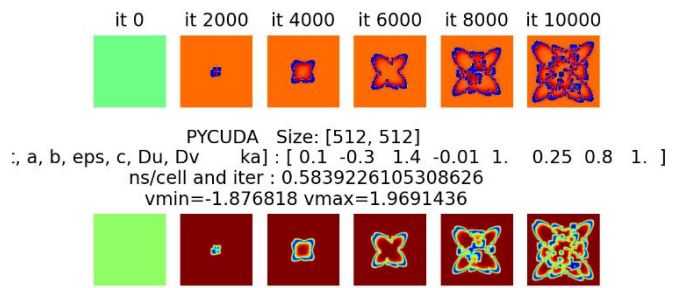

Fig. 10. A simulation of RD-CNN for the a specific case of "slow-growth"

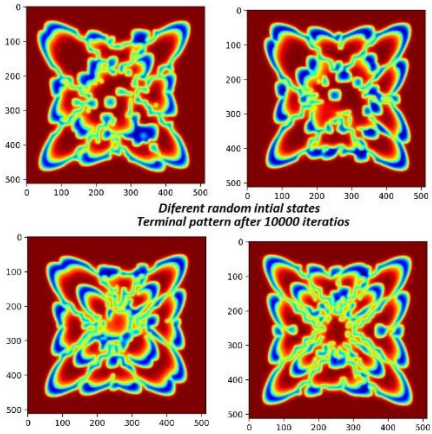

Fig. 11. Four different "slow-growth" patterns emergin from different mid-random square initial states.

### B. Image processing applications (edge detection)

Another interesting application of emergent dynamics, with potential relevance in artificial intelligence, is the detection of features from initial state images. Runing a fast simulator allow fast identification of meaningful processing behaviors, as shown in Figure 12 where some sets of *(a,b)* parameters can be identified in relationship with meaningful behaviors.

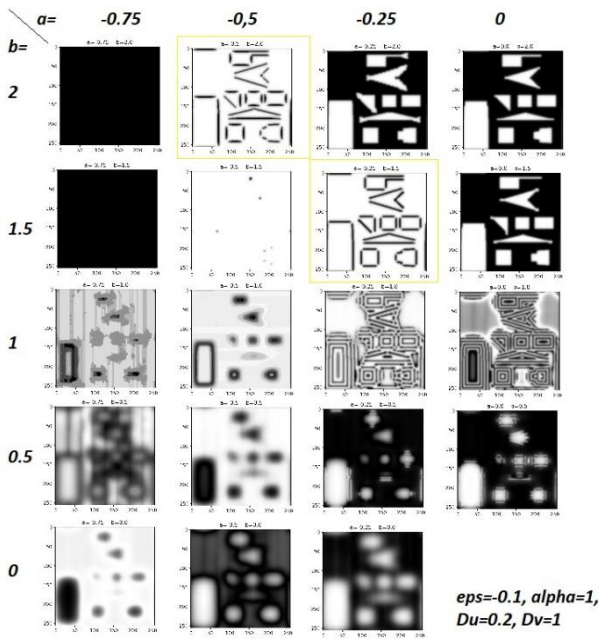

Fig. 12. Parameter search for meaningful image processing bahaviors.

Other set of parameters, revealing meaningful behaviors, relevant for the edge detection task were already exposed in the simulations shown in Figures 1 and 2.

## IV. CONCLUDING REMARKS

A freely available modelling and simulation framework (MSF) for fast simulation of cellular nonlinear networks is proposed [16]. It can be easily extended to any FDTD (finite-difference time-domain) model while it was designed for reproducible research and education using the readily available resources from Google Collaboratory. Four variants of implementations using four Python libraries were considered. The fastest version relies on PyCUDA library, capable to achieve as much as 14000 Mcells/second speed using the best GPU platform available in the cloud platform (Tesla P100-PCIE). As a comparison, the freely available Ready package [9] which can be installed on a personal computer ensures no more than 2300 Mcells / second for a grid size of 1024x1024 cells. Of course, the performance here was limited by the local GPU, in this case a Geforce GTX950. Installing a better GPU on a local computer to run Ready or other similar simulator can be rather costly in comparison to the advantage of our cloud-computing based MSF which can freely exploit some better available GPUs. Other important advantage of our solution is related to the possibility to easily embed basic functionalities in more sophisticated applications using the widely available Python libraries. For instance, our further research will focus on combining the MSF (patterns generated by it) with deep learning solutions to classify automatically various types of emergent behaviors, thus providing some more accurate alternative solution to the local activity theory for identification of interesting dynamic behaviors while searching the parameters space and performing rapid simulations of the implemented cellular nonlinear model.

## REFERENCES

[1] L. O. Chua, "CNN: A paradigm for complexity," in Visions of Nonlinear Science in the 21st Century. Eds. Singapore: World Scientific, 1999, vol. 26.

[2] A. Taflove; S.C. Hagness, Computational Electrodynamics: The Finite-Difference Time-Domain Method; Artech House Publishers, 2005.

[3] R. Dogaru, Systematic design for emergence in cellular nonlinear networks – with applications in natural computing and signal processing, Springer-Verlag, Berlin Heidelberg, 2008

[4] R. Dogaru, Universality and Emergent Computation in Cellular Neural Networks, World Scientific, Singapore, 2003.

[5] L. O. Chua, "Local activity is the origin of complexity", Int. J. Bifurcation Chaos, vol. 15, no. 11, pp. 3435-3456, 2005.

[6] M. Bucurica, I. Dogaru and R. Dogaru, "Improving computational efficiency for implementing a sound propagation simulation environment using Python and GPU," 2016 8th International Conference on Electronics, Computers and Artificial Intelligence (ECAI), Ploiesti, 2016, pp. 1-4

[7] H. Zhu, X. Xu, G. Huang, Z. Qin and B. Wen, "An Efficient Graphics Processing Unit Scheme for Complex Geometry Simulations Using the Lattice Boltzmann Method," in IEEE Access, vol. 8, pp. 185158-185168, 2020.

[8] P. F. Baumeister, T. Hater, J. Kraus, D. Pleiter and P. Wahl, "A Performance Model for GPU-Accelerated FDTD Applications," 2015 IEEE 22nd International Conference on High Performance Computing (HiPC), Bengaluru, India, 2015, pp. 185-193.

[9] Tim Hutton, Robert Munafo, Andrew Trevorrow, Tom Rokicki, Dan Wills, "Ready, a cross-platform implementation of various reaction-diffusion systems." https://github.com/GollyGang/ready

[10] R. J. LeVeque, I. M. Mitchell and V. Stodden, "Reproducible research for scientific computing: Tools and strategies for changing the culture," in Computing in Science & Engineering, vol. 14, no. 4, pp. 13-17, July-Aug. 2012.

[11] T. Kluyver, B. Ragan-Kelley, F. Perez, B. Granger, M. Bussonnier, J. Frederic, K. Kelley, J. Hamrick, J. Grout, S. Corlay, P. Ivanov, D. Avila, S. Abdalla, and C Willing, "Jupyter notebooks – a publishing format for reproducible computational workflows", In F. Loizides and B. Schmidt, editors, Positioning and Power in Academic Publishing: Players, Agents and Agendas, pages 87 – 90. IOS Press, 2016.

[12] Nelson, M. J. & Hoover, A. K. Notes on Using Google Colaboratory in AI Education. In Proceedings of the 2020 ACM Conference on Innovation and Technology in Computer Science Education, 533–534, 2020.

[13] Radu Dogaru, Ioana Dogaru, "High Productivity Cellular Neural Network Implementation on GPUs using Python", published in Proceedings of the Workshop in Information Technology and Bionics (Symposium in Memory of Tamás Roska), Budapest, 23–24 June 2015, ISBN 978-963-89880-3-4, pp. 23-27.

[14] R. Dogaru,"Applications of Emergent Computation in Reaction-Diffusion CNNs for Image Processing," Control Systems and Computer Science (CSCS), 2013 19th International Conference on , vol., no., pp.370,377, 29-31 May 2013.

[15] R. Dogaru and I. Dogaru, "A Low Cost High Performance Computing Platform for Cellular Nonlinear Networks Using Python for CUDA," 2015 20th International Conference on Control Systems and Computer Science, Bucharest, Romania, 2015, pp. 593-598.

[16] R. Dogaru, I. Dogaru, code support for this article: https://github.com/radu-dogaru/fast-fhn-rd-cnn-simulators

[17] Kaggle Documentation. Online: https://www.kaggle.com/docs (2020).

[18] Andreas Klöckner, Nicolas Pinto, Yunsup Lee, Bryan Catanzaro, Paul Ivanov, Ahmed Fasih, PyCUDA and PyOpenCL: A scripting-based approach to GPU run-time code generation, Parallel Computing, Volume 38, Issue 3, March 2012, Pages 157-174.

[19] https://numba.pydata.org/numba-doc/dev/cuda/index.html

[20] Ryosuke Okuta, Yuya Unno, Daisuke Nishino, Shohei Hido and Crissman Loomis. CuPy: A NumPy-Compatible Library for NVIDIA GPU Calculations. Proceedings of Workshop on Machine Learning Systems (LearningSys) in The Thirty-first Annual Conference on Neural Information Processing Systems (NIPS), (2017).